\documentclass[onecollarge,natbib]{svjour2}
\bibpunct{[}{]}{;}{n}{}{,} 
\smartqed  
\usepackage{graphicx}
\usepackage{amsmath}
\journalname{Few-Body Systems (EFB22)}
\begin{document}

\title{Ab initio NCSM/RGM for
three-body cluster systems and application to $^4$He+n+n
}


\author{C. Romero-Redondo \and
P. Navr\'atil \and
S. Quaglioni   \and
G. Hupin 
}


\institute{C. Romero-Redondo \at
              TRIUMF, 4004 Wesbrook Mall, Vancouver, BC V6T 2A3, Canada \\
              Tel.: +1-604-222-1047 ext. 6447 \\
              Fax:  +1-604-222-1074\\
              \email{cromeroredondo@triumf.ca}           
\and P. Navr\'atil \at
              TRIUMF, 4004 Wesbrook Mall, Vancouver, BC V6T 2A3, Canada 
           \and
           S. Quaglioni \and G. Hupin \at
              Lawrence Livermore National Laboratory, P.O. Box 808, L-414, Livermore, California 94551, USA
}

\date{Received: date / Accepted: date}

\maketitle

\begin{abstract}
We introduce an extension of the $ab~initio$
 no-core shell model/resonating group method (NCSM/RGM) in order to
describe three-body
cluster states.  We present results for the $^6$He ground state within a $^4$He+n+n cluster
basis as well as first results for the phase shifts 
 of different channels of the $^4$He+n+n system which provide information about
low-lying resonances of this nucleus. 
\end{abstract}

\section{Introduction}
\label{intro}
The $ab~initio$ NCSM/RGM was
presented in \cite{QuaPRL08,QuaPRC09} as a promising technique that is able to
treat  both structure and reactions in light nuclear systems. This approach combines  a microscopic cluster technique with the use of realistic interactions and a consistent $ab~initio$ description of the nucleon clusters.

The method has been introduced in detail for two-body cluster bases and has
been shown to work efficiently in different systems \cite{QuaPRL08,QuaPRC09,NavPRC11,NavPRL12}. However, there are
many interesting systems that have a three-body cluster structure and
therefore can not be successfully studied with a two-body cluster approach.

The extension of the NCSM/RGM approach to properly describe three-body cluster
states is essential for the study of nuclear systems that present
such configuration. This type of systems appear, $e.g$,  in
structure problems of
two-nucleon halo nuclei such as $^6$He and $^{11}$Li, resonant systems like $^5$H or 
transfer reactions with
three fragments in their final states like $^3$H($^3$H,2n)$^4$He or
$^3$He($^3$He,2p)$^4$He. 

Recently, we introduced three-body cluster
configurations into the method and presented the first results for the $^6$He ground state
\cite{Quaglioni:2013kma}. Here we present these results as well as first results for the
continuum states of $^6$He within a $^4$He+n+n basis.

\section{Formalism}
\label{sec:1}

The extension of the
NCSM/RGM approach to properly describe three-cluster configurations requires
to expand the many-body wave function 
over a basis $|\Phi_{\nu xy}^{J^{\pi}T}\rangle$ of three-body cluster channel
states built from the NCSM wave function of each of the three clusters, 

\begin{equation}
|\Psi^{J^{\pi}T}\rangle=\sum_{\nu}
\int dxx^2\int dy y^2 G_{\nu}^{J^{\pi}T}(x,y)\hat{A}_{\nu}
\mbox{{\boldmath $|\Phi_{\nu xy}^{J^{\pi}T}\rangle 
$}}
\label{eq1}
\end{equation}

\begin{eqnarray}
         |\Phi^{J^\pi T}_{\nu x y} \rangle  = & 
        \Big[\Big(|A-a_{23}~\alpha_1I_1^{\pi_1}T_1\rangle 
        \left (|a_2\, \alpha_2 I_2^{\pi_2} T_2\rangle |a_3\, \alpha_3 I_3^{\pi_3}T_3\rangle \right)^{(s_{23}T_{23})}\Big)^{(ST)} 
       \nonumber \\ & \left(Y_{\ell_x}(\hat{\eta}_{a_2-a_3})Y_{\ell_y}(\hat{\eta}_{A-a_{23}})\right)^{(L)}\Big]^{(J^{\pi}T)} 
         \times \frac{\delta(x-\eta_{a_2-a_3})}{x\eta_{a_2-a_3}} \frac{\delta(y-\eta_{A-a_{23}})}{y\eta_{A-a_{23}}}\,,
        \label{eq:3bchannel}    
\end{eqnarray}

where $\eta_{1,23}$
is the relative vector proportional to the displacement between the center of mass (c.m.) of the first cluster and that of the residual two fragments,
 and $\eta_{23}$
is the relative coordinate proportional to the distance between the centers of mass of cluster
2 and 3.
In eq. (\ref{eq1}), $G_{\nu}^{J^{\pi}T}(x,y)$ are the relative motion wave functions and represent
the unknowns of the problem and $\hat{A}_{\nu}$ is the intercluster 
antisymmetrizer.  

Projecting the microscopic $A$-body Schr\"odinger equation onto the basis states $\hat {\mathcal A}_\nu\, |\Phi^{J^\pi T}_{\nu x y} \rangle$, the many-body problem can be mapped onto the system of coupled-channel integral-differential equations
\begin{eqnarray}
        \sum_\nu \iint \!\!dx \, dy \, x^2 y^2 \Big [ {\mathcal H}^{J^\pi T}_{\nu^\prime\nu}(x^\prime,y^\prime,x,y) 
         - E \, {\mathcal N}^{J^\pi T}_{\nu^\prime\nu}(x^\prime,y^\prime,x,y) \Big] G_{\nu}^{J^\pi T}(x,y) = 0,\label{eq:3beq1} 
\end{eqnarray}
where $E$ is the total energy of the system in the c.m.\ frame and
\begin{eqnarray}
        {\mathcal H}^{J^\pi T}_{\nu^\prime\nu}(x^\prime,y^\prime,x,y) & = 
                \left\langle\Phi^{J^\pi T}_{\nu^\prime x^\prime y^\prime} \right| \hat {\mathcal A}_{\nu^\prime} H \hat {\mathcal A}_\nu \left | \Phi^{J^\pi T}_{\nu x y} \right\rangle\,, \label{eq:Hkernel} \\
        {\mathcal N}^{J^\pi T}_{\nu^\prime\nu}(x^\prime,y^\prime,x,y) & = 
                \left\langle\Phi^{J^\pi T}_{\nu^\prime x^\prime y^\prime} \right| \hat{\mathcal A}_{\nu^\prime}  \hat {\mathcal A}_\nu \left | \Phi^{J^\pi T}_{\nu x y} \right\rangle   \label{eq:Nkernel}
\end{eqnarray}
are integration kernels given respectively by the Hamiltonian and overlap (or norm) matrix elements over the antisymmetrized basis states.  Finally, $H$ is the intrinsic $A$-body Hamiltonian.

In order  to solve the Schr\"odinger equations (\ref{eq:3beq1}) we
orthogonalize them and
transform to the  hyperspherical harmonics (HH) basis to obtain a set of 
non-local integral-differential equations in the hyper-radial coordinate,

\begin{eqnarray}
        \sum_{K\nu}\int d\rho \rho^5 \bar{\cal H}_{\nu'\nu}^{K'K}(\rho',\rho) \frac{u^{J^{\pi}T}_{K\nu}(\rho)}{\rho^{5/2}} 
        = E \frac{u^{J^{\pi}T}_{K^\prime\nu^\prime}(\rho^\prime)}{\rho^{\prime\,5/2}}\, ,
        \label{RGMrho}
\end{eqnarray}
which is finally solved using the microscopic R-matrix method on a Lagrange 
mesh. The details of the procedure  can be found in 
\cite{Quaglioni:2013kma}.

At present, we have completed the development of the formalism for the
treatment of three-cluster systems formed by two separate nucleons in
relative motion with respect to a nucleus of mass number A$-2$.

\section{Application to $^4$He+n+n.}
\label{sec:2}

It is well known that
$^6$He is the lightest Borromean nucleus~\cite{Tanihata:1995yv, PhysRevLett.55.2676}, formed by an $^4$He core and two halo neutrons. It is, therefore, 
 an ideal first candidate to be studied within this approach.
In the present calculations, we describe the $^4$He core only by its g.s.
 wave function, ignoring its excited states. 
This is the only limitation in the model space used. 

We used similarity-renormalization-group
 (SRG) \cite{SRG,SRG2} evolved potentials obtained from the
 chiral N$^3$LO NN interaction \cite{NN}
with $\Lambda$ = 1.5 fm$^{-1}$. The set of equations (\ref{RGMrho}) are solved for different
channels using both bound and continuum asymptotic conditions. We find only one bound state, which
appears  in the 
$J^{\pi}T=0^+1$ channel and corresponds to the $^6$He ground state.

\paragraph{Ground state}

\begin{table}[t]
\caption{Ground-state energies of the $^4$He and $^6$He nuclei. Extrapolations were performed with an exponential fit.}
\centering
\label{tab:a}
\begin{tabular}{lccc}
\hline\noalign{\smallskip}
   Approach
  &
  &  E$_{g.s}$($^4$He)
  &  E$_{g.s}$($^6$He)  \\[3pt]
\tableheadseprule\noalign{\smallskip}
NCSM/RGM & ($N_{\rm max}$=12) & $-28.22$ MeV     & $-28.70$ MeV \\
NCSM & ($N_{\rm max}$=12) & $-28.23$ MeV  & $-29.75$ MeV\\
NCSM & (extrapolated) & $-28.23(1)$ MeV  & $-29.84(4)$ MeV\\
\noalign{\smallskip}\hline
\end{tabular}
\end{table}

The results for the g.s. energy of $^6$He within a $^4$He(g.s.)+n+n
cluster basis and $N_{\rm max}$ = 12, $\hbar\Omega$
 = 14 MeV harmonic oscillator model space are compared to
NCSM calculations in table \ref{tab:a}. 
At $N_{\rm max} \sim$ 12 the binding energy
calculations are close to convergence in both NCSM/RGM and NCSM approaches. The
observed difference of approximately 1 MeV is due to the excitations of the
$^4$He core, included only in
the NCSM at present. Therefore, it gives a measure of the polarization effects
of the core. The inclusion of the excitations of the core will be achieved in 
a future work through the use of the no-core shell model with continuum 
approach (NCSMC) \cite{PhysRevLett.110.022505,PhysRevC.87.034326}, which couples
 the present three-cluster wave functions with NCSM eigenstates of the six-body system. 

\begin{figure}
\centering
  \includegraphics[height=6cm]{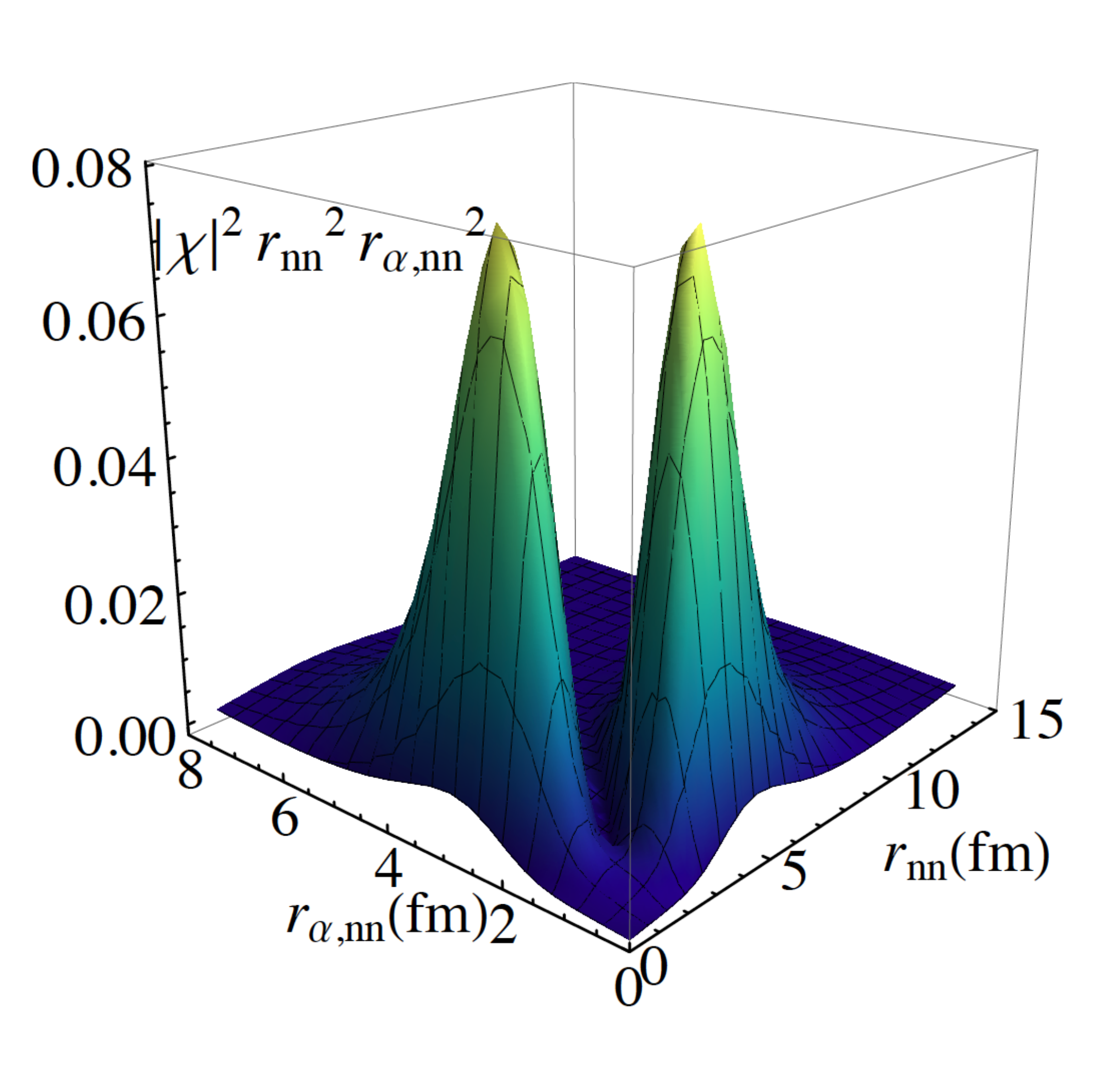}
\hspace{0.8cm}
  \includegraphics[height=5.5cm]{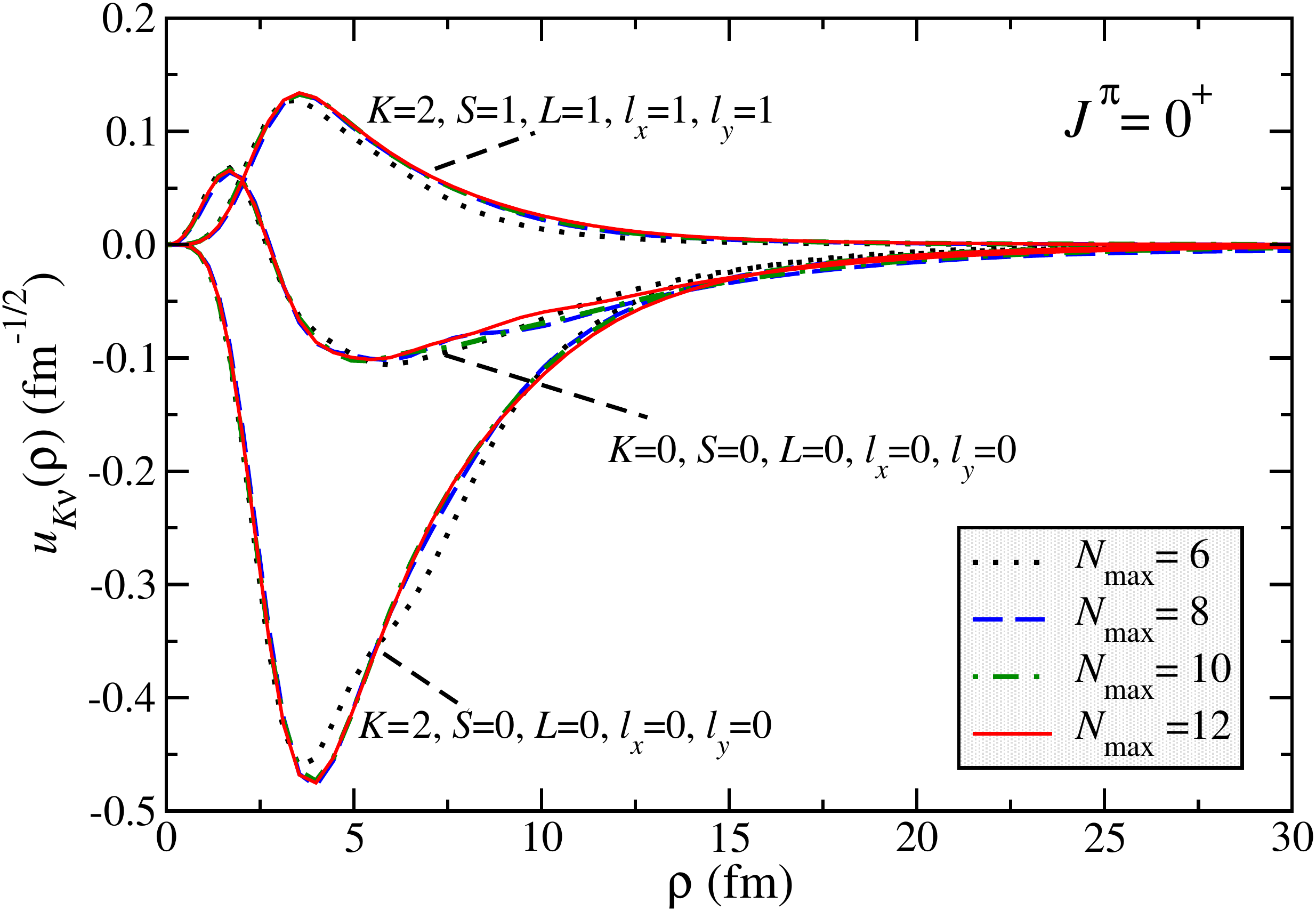}
\caption{The figure in the left shows the probability distribution of the main component of the $^4$He+$n$+$n$
relative motion wave function for the $J^\pi T=0^+1$ ground state, $r_{nn}$ and $r_{\alpha,nn}$ are
the distances between the two neutrons and between the $\alpha$ particle and center of mass of the 
two neutrons, respectively. 
In the right, the three main components of the radial part of the $^6$He g.s.\ wave
functions $u_{K \nu}(\rho)$ for $N_{\rm max}$=6,8,10, and 12. }
\label{fig:1}       
\end{figure}

 Contrary to the NCSM, in the NCSM/RGM the $^4$He(g.s.)+n+n
wave functions present the appropriate asymptotic behavior. The main components
of the radial part of the $^6$He g.s.\ wave
function $u_{K \nu}(\rho)$ can be seen in fig. (\ref{fig:1}) for different sizes of the model
space demostrating large extension of the system. In the left part of 
the figure, the probability distribution of the main component
of the wave function is shown, featuring two characteristic peaks which correspond 
to the di-neutron and cigar
configurations.

A thorough study of the converge of the results with respect to
different parameters of the calculation was presented in 
\cite{Quaglioni:2013kma}, showing good convergence and stability.

\paragraph{Continuum states}
The use of three-cluster dynamics is  essential for describing 
$^6$He states in the continuum. Therefore, this formalism is ideal for such
study.
Using continuum asymptotic conditions, we solved the set of equations (\ref{RGMrho})
in order to obtain the low-energy phase shifts for the
$J^{\pi}=0^+, 1^-, 1^+$ and $ 2^+$ channels in the continuum. 

\begin{figure}
\centering
  \includegraphics[height=6cm]{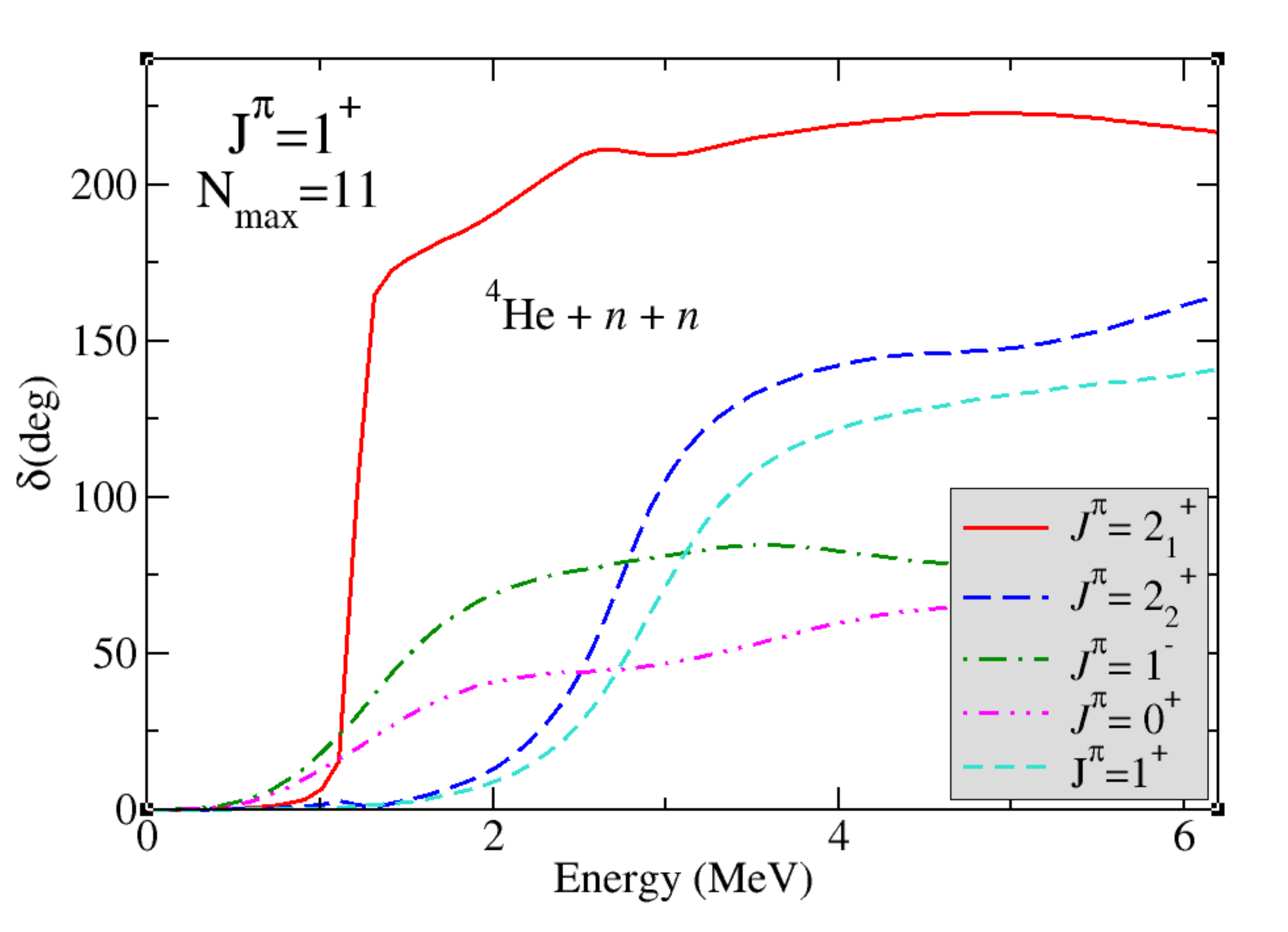}
\caption{Preliminary diagonal phase shifts for $^4$He for different $J^\pi$ channels.}
\label{fig:2}       
\end{figure}

In our preliminary results, we obtain the experimentally well-known $2^+_1$ resonance as well 
as a second low-lying
$2^+$ resonance recently measured at Ganil \cite{Mougeot:2012aq}. 
A resonance is also found in the $1^+$ channel while no low-lying resonances are present in
the $0^+$ or $1^-$ channels.
In fig. \ref{fig:2}
some of the preliminary phase shifts for different channels are shown. 
Results for bigger model spaces and a study of their stability respect to the parameters 
in the formalism are 
presently being calculated and will be
presented elsewhere.

%

\section{Conclusions}
In this work, we present an
extension of the NCSM/RGM which includes three-body dynamics in the formalism. 
This new feature permits us to study  a new range of
systems that present three-body configurations 
.  In particular,
we presented results for both bound and continuum states of  $^6$He studied within 
a basis of $^4$He+n+n.  The obtained wave functions feature an 
appropriate asymptotic behavior, contrary to bound-state $ab~initio$ methods such as 
the NCSM.

\begin{acknowledgements}
Computing support for this work came from the LLNL institutional Computing Grand Challenge
program and from an INCITE Award on the Titan supercomputer of the Oak Ridge Leadership Computing Facility (OLCF) at ORNL. Prepared in part by LLNL under Contract DE-AC52-07NA27344. Support from the U.S. DOE/SC/NP (Work Proposal No. SCW1158) and
NSERC Grant No. 401945-2011
is acknowledged. TRIUMF receives funding via a contribution through the
Canadian National Research Council.
\end{acknowledgements}



\end{document}